\documentclass[12pt]{article}
\usepackage{graphics}
\begin{document}

\thispagestyle{empty}

\begin{center}
{\large\bf Local Bursts Model of CMB Temperature Fluctuations: Scattering in
Resonance Lines of Primordial Hydrogen and Helium}
\end{center}

\begin{center}
V.K. Dubrovich$^1$, S.I. Grachev$^2$
\end{center}

\begin{center}

$^1$Special Astrophysical Observatory, St. Petersburg Branch, Russian Academy
of Sciences, St. Petersburg, 196140 Russia\\
$^2$Sobolev Astronomical Institute, St. Petersburg State University,
Universitetskii pr. 28, St. Petersburg, 198504 Russia\\
\end{center}

{\small Within the framework of a flat cosmological model a propagation of an
instantaneous burst of isotropic radiation is considered from the moment of
its beginning at some initial redshift $z_0$ to the moment of its registration
now (at $z=0$). We take into account Thomson scattering by free electrons and
scattering in L$_\alpha$ and L$_\beta$ lines of primordial hydrogen and in lines
1s2 -- 1s2p, 1s3p ($^1$S -- $^1$P*) of HeI. It is shown that
relative amplitude of spectrum distortions caused by scattering in these lines
may be from $10^3$ to $10^4$ times greater than maximum possible amplitude due
to scattering in subordinate lines considered in our previous paper (Dubrovich,
Grachev, 2015). In a linear approximation on the optical thickness in the lines
the profiles of distortions in resonance lines turn out to be purely in
absorption and do not depend on both direction and distance to the burst center
in contradistinction to the profiles in subordinate lines. The profiles contain
the jumps on the frequencies corresponding to appearance of a source (a burst)
at a given redshift $z = z_0$. For $z_0=5000$ the jumps in hydrogen L$_\alpha$
and L$_\beta$ lines are on the frequencies 493 and 584 GHz respectively and the
jumps in the two HeI lines are on 855 and 930 GHz for $z_0=6000$. Relative
magnitude of jumps amounts to $10^{-4}$ --- $3\cdot 10^{-3}$.

\bigskip

\noindent {\it Key words:} cosmology, early Universe, cosmological
recombination, radiative transfer, Thomson scattering, line scattering
}

\begin{center}
{\bf Introduction}
\end{center}

New results of the PLANCK mission (see Adam et al., 2015a) greatly improve
exactness of our knowlege about the power spectrum of primordial spatial
fluctuations of the matter density in the early Universe, about peculiarities
of primordial matter recombination dynamics, about some global fundamental
parameters of the Universe and about CMB polarization. Essential progress
is achieved also in the CMB spectroscopy. However all these important successes
do not cancel further more detailed and deep investigations of CMB properties.

In particular the medium resolution spectroscopy of separate objects on the CMB
brightness map seems to be very important. Novelty here is in turning from
investigation of statistical CMB properties to searching and learning local
phenomena and ``protoobjects''. The last ones may be somewhat rare events
having practically no effect on the average statistical CMB parameters. But
they can carry some information about completely new physical laws. So for
example one can expect an existence of primordial black holes of great mass
and new local forms of matter and fields (see, for example, Dubrovich (2003),
Grachev and Dubrovich (2011), Dubrovich and Glazirin (2012)).

Besides of more or less probable but still hypothetical objects there
evidently exists the whole class of local sources in the early Universe which
can be learned separately. These are the same standard primordial CMB 
temperature fluctuations thoroughly learned now statistically. In fact we deal
with some spatial domains where temperature increase or decrease takes place for
some reasons or other. It is very important that besides spatial apartness of
these regions their temperature deviation is also nonstationary. Depending on
a mechanism of a given inhomogeneity formation a typical time of its
development and damping can be different. With regard to such events one can
speak about local-bursts model of fluctuations.

The present work is a continuation of our previous paper (Dubrovich and Grachev,
2015) devoted to calculations of radiation field evolution due to Thomson
scattering and scattering in subordinate lines  H$_\alpha$, H$_\beta$,
P$_\alpha$ and P$_\beta$ of primordial hydrogen in a homogeneous expanding and
recombining Universe in assumption that radiation arises instantaneously at
some redshift $z_0$. Now we calculate spectral distortions arising from
scattering in HI and HeI lines of Lyman series. Several distanses from the
burst center and several initial redshifts $z_0$ are considered. The moments
$z_0$ are
selected in such a way that optical thicknesses of the Universe in these
lines were less than unity. Polarization of radiation is neglected. Scattering
both Thomson and in lines is assumed to be isotropic. The absorption coefficient
profile is approximated by delta-function in the comoving frame of reference
i.e. we neglect own line broadening in comparison with the broadening due to
space expansion. It is assumed also that the burst radiation do not affect on
electron concentration and on occupation numbers of atom levels which are
calculated beforehand by using Boltzmann and Saha equations because conditions
in considered range of redshifts ($z=3500 - 6000$) are close to the local
thermodynamic equilibrium (LTE).

\begin{center}
{\bf The main equations and relations}
\end{center}

We are interested in distortions in a source (burst) spectrum which may arise
as a result of scattering in HI and HeII resonance lines in the epoch when
optical thickness of the Universe in these lines was sufficiently small ($<1$).
In Sobolev approximation (a narrow line approximation) an optical thickness in
a line arising in a transition $1\leftrightarrow k$, is given by (see e.g.
Dubrovich and Grachev, 2004)
\begin{equation}
\tau_{1k}(z)=\frac{hc}{4\pi}\frac{n_1(z)B_{1k}}{H(z)}\left[1-\frac{n_k(z)g_1}
{n_1(z)g_k}\right],
\label{tauik}
\end{equation}
where $h$ is the Planck constant, $c$ is the speed of light, $n_1$ an $n_k$ are
occupation numbers of the levels, $g_1$ and $g_k$ are statistical wheights of
the levels, $B_{1k}$ is the Einstein coefficient for radiation absorption in the
transition $1\rightarrow k$, $H(z)$ is the Hubble factor.

Let us find concentrations of atoms and ions by using Saha formulae. Let
$N_{\rm H}$ and $N_{\rm He}$ are total concentrations of atoms and ions for
hydrogen and helium correspondingly, $f_{\rm He}\equiv N_{\rm He}/N_{\rm H}$ is
the helium abundance in number of atoms. Further let $N_{\rm H^0}$ and
$N_{\rm H^+}$ of hydrogen neutral atoms and ions respectively, $N_{\rm He^0}$,
$N_{\rm He^+}$ and $N_{\rm He^{++}}$ are concentrations of neutral, singly
ionized and doubly ionized helium atoms. We will measure all cocentrations in
units $N_{\rm H}$: $x_{\rm H^0}=N_{\rm H^0}/N_{\rm H}$,
$x_{\rm H^+}=N_{\rm H^+}/N_{\rm H}$, $x_{\rm He^0}=N_{\rm He^0}/N_{\rm H}$,
$x_{\rm He^+}=N_{\rm He^+}/N_{\rm H}$, $x_{\rm He^{++}}=
N_{\rm He^{++}}/N_{\rm H}$ and $x_{\rm e}=n_{\rm e}/N_{\rm H}$, where $n_{\rm e}$
is the electron concentration. Then Saha equations and charge conservation
equation are written as
\begin{equation}
x_{\rm e}\frac{x_{\rm H^+}}{1-x_{\rm H^+}}=\frac{2U({\rm H^+})}{U({\rm H^0})}
\frac{F(T_{\rm e})}{N_{\rm H}}\exp(-\chi_{\rm H}/kT_{\rm e}),
\end{equation}
\begin{equation}
x_{\rm e}\frac{x_{\rm He^+}}{f_{\rm He}-x_{\rm He^+}-x_{\rm He^{++}}}=
\frac{2U({\rm He^+})}{U({\rm He^0})}
\frac{F(T_{\rm e})}{N_{\rm H}}\exp(-\chi_{\rm He}/kT_{\rm e}),
\end{equation}
\begin{equation}
x_{\rm e}\frac{x_{\rm He^{++}}}{x_{\rm He^+}}=
\frac{2U({\rm He^{++}})}{U({\rm He^+})}
\frac{F(T_{\rm e})}{N_{\rm H}}\exp(-\chi_{\rm He^+}/kT_{\rm e}),
\end{equation}
\begin{equation}
x_{\rm e}=x_{\rm H^+}+x_{\rm He^+}+2x_{\rm He^{++}},
\end{equation}
where $F(T_{\rm e})=(2\pi m_{\rm e}kT_{\rm e})^{3/2}h^{-3}$, $U$ are partition
functions ($U({\rm H^0})=2$, $U({\rm H^+})=1$, $U({\rm He^0})=1$,
$U({\rm He^+})=2$, $U({\rm He^{++}})=1$), $\chi$ are ionization potentials,
$k$ is the Boltzmann constant, $m_{\rm e}$ is the electron mass, $T_{\rm e}$ is
the electron temperature. For the considered range of redshifts one may adopt
that the electron temperature coinsides with the temperature of background
blackbody radiation: $T_{\rm e}(z)=T(z)=T_0(1+z)$, where $T_0$ is the
present-day CMB temperature.

By using Boltzmann equation for occupation numbers and taking into account that
the overwhelming part of atoms is in the ground state, i.e.
$n_{\rm 1,H^0}=N_{\rm H^0}=x_{\rm H^0}N_{\rm H}$, $n_{\rm 1,He^0}=
x_{\rm He^0}N_{\rm H}$, one can rewrite eq. (\ref{tauik}) for HI and HeI lines
in the form
\begin{equation}
\tau_{1k}(z)=\frac{c^3}{8\pi\nu_{1k}^3}\frac{x_0(z)N_{\rm H}(z)}{H(z)}
\frac{g_k}{g_1}A_{k1}\left[1-{\rm e}^{-h\nu_{1k}/kT(z)}\right],
\label{tauik1}
\end{equation}
where $h\nu_{1k}$ is the transition energy, $A_{k1}$ is the Einstein
coefficient of spontaneous transitions, $x_0=x_{\rm H^0}$ or $x_{\rm He^0}$ for
HI or HeI respectively.

As for the other parameters (except for $T_0$) appearing in the problem they
enter in particular in the Hubble factor
\begin{equation}
H(z)=H_0\sqrt{\Omega_{\Lambda}+(1-\Omega)(1+z)^2+\Omega_{M}(1+z)^3+\Omega_{rel}
(1+z)^4},
\end{equation}
where $H_0=2.4306\cdot 10^{-18}h_0$ с$^{-1}$, $h_0$ is the Hubble constant in
the units 75 km/(s$\cdot$Mpc); $\Omega_{\rm M}$, $\Omega_{\Lambda}$ and
$\Omega_{\rm rel}$ are the ratios of matter, dark energy and relativistic
particles (radiation, massless neutrino) densities to the crytical density now;
$\Omega=\Omega_{\rm M}+\Omega_{\Lambda}+\Omega_{\rm rel}$,
$\Omega_{\rm rel}=\rho^0_R(1+f_n)/\rho_c$, $\rho^0_R=a_RT_0^4/c^2$ is the
radiation mass density now, $f_n$ is the part of relativistic (massless)
neutrino (usually $f_n=0.68$). For the flat model of the Universe $\Omega=1$,
and then $\Omega_{\rm M}=1-\Omega_{\Lambda}-\Omega_{\rm rel}$.

Moreover the total concentration of hydrogen atoms and ions
\begin{equation}
N_{\rm H}(z)=N_{\rm H}^0(1+z)^3,\quad  N_{\rm H}^0=0.63144\cdot10^{-5}X
\Omega_{\rm B}h_0^2 \,\mbox{см}^{-3},
\end{equation}
enter the equations. Here $\Omega_{\rm B}$ is the ratio of baryon density to
the crytical density now, $X$ is the hydrogen abundance (by mass).
As the base we use the following values of parameters: $\Omega=1$,
$\Omega_\Lambda=0.7$, $\Omega_{\rm B}=0.04$, $T_0=2.728$ K, $X=0.76$,
$\Omega_{\rm rel}=0.85\cdot 10^{-4}$, the Hubble constant $H_0=70$
km/(s$\cdot$Mpc).

Figure 1 shows optical thicknesses in hydrogen L$_\alpha$ and L$_\beta$ lines
and in HeI lines 1s2-1s2p ($^1$S -- $^1$P*) and 1s2-1s3p ($^1$S -- $^1$P*). We
choose such a range of redshifts in order that the optical thicknesses
($\tau_{1k}$)
of the Universe in these lines were less than unity. For hydrogen $\tau_{12}<1$
for $z>3060$ and $\tau_{13}<1$ for $z>2750$ which corresponds to the
frequency ranges $\nu<805$ and $\nu<1062$ GHz, where $\nu=\nu_{1k}/(1+z)$,
$\nu_{1k}$ is the laboratory frequency of transition. For the two considered
HeI lines the corresponding ranges of frequencies are $\nu<1120$ and
$\nu<1310$ GHz.

Besides optical thicknesses one must know albedo of a single scattering
$\lambda_{1k}(z)$ in the considered lines. When calculating albedo we take
into account spontaneous and induced transitions due to blackbody background
radiation with the temperature $T(z)$:
\begin{equation}
\lambda_{1k}(z)=R_{k1}(z)/[\sum_{i'=1,i'\neq k}^{\infty}R_{ki'}(z)+
R_{kc}(z)],
\label{lik}
\end{equation}
where
\begin{equation}
R_{ki}(z)=\frac{g_i}{g_k}\frac{A_{ik}}{\exp[h\nu_{ki}/kT(z)]-1},\quad k<i,
\label{Rkid}
\end{equation}
are the coefficients of transitions upwards and
\begin{equation}
R_{ki}(z)=\frac{A_{ki}}{1-\exp[-h\nu_{ik}/kT(z)]},\quad k>i,
\label{Rkiu}
\end{equation}
are the coefficients of transitions downwards, $R_{kc}$ are the coefficients
of transitions due to ionization by blackbody radiation, $A_{ki}$ are
Einstein coefficients of spontaneous transitions. For hydrogen we calculate
albedo by using 60-levels atom model (see Grachev and Dubrovich, 1991) and for
HeI we take into account transitions 1s2p, 1s3p -- 1sks ($^1$P* -- $^1$S), 1skd
($^1$P* -- $^1$D) for k$\leq 10$ (the data from the base NIST: Ralchenko et
al., 2011). For HeI photoionization crossections from the states k$^1$P* for k=2
and 3 we use approximate formula $\sigma_{kc}(\nu)=\sigma_0(\nu_{kc}/\nu)^3$
where $\sigma_0$ and $\nu_{kc}$ are the threshold values of the crossection and
frequency of ionization. According to Hummer and Storey (1998)
$\sigma_0=1.35\cdot 10^{-17}$ and $2.70\cdot 10^{-17}$ cm$^2$ for k = 2 and 3
respectively. Figure 2 shows albedos of a single scattering in considered lines.

To solve the problem of radiation transfer taking into account both Thomson
scattering and scattering in resonance lines we use the same method as in our
previous paper (Dubrovich and Grachev, 2015) devoted to subordinate hydrogen
lines. The only additional condition is a limitation on the value of the
optical thickness in resonance lines. The thing is that in contradiction to
subordinate lines which are always optically thin resonance lines can become
optically very thick starting with some moment of the Universe evolution.
In this case the calculations must be fulfilled with a proper account of
multiple scatterings which is much more difficult. So we consider the period
of evolution corresponding to sufficiently large redshifts when the optical
thickness of the Universe in HI and HeI resonance lines was less than unity.
To the first order in line optical thickness we have obtained (Dubrovich and
Grachev, 2015, eq. (38)) the following relative energy distribution in a burst
spectrum in the present-day epoch ($z=0$, $u=u_0$, $\eta=\eta_0$):
\begin{equation}
\frac{n-n^0}{n^0}=-\tau_l\left\{1-\frac{\lambda_l}{n^0}\left[
e^{u_l-u_0}s^0(r'_l,u_l)+\int_{u_l}^{u_0} s^0(r',u')e^{u'-u_0}du'
\right]\right\},
\label{nexp0}
\end{equation}
where $n$ is the radiation intensity (in average photon occupation numbers),
$n^0$ and $s^0$ are radiation intensity and source function without account of
line scattering (but with a proper account of multiple scattering on free
electrons), $\tau_l=\tau_{1k}(z_l)$, $u_l=u(z_l)$,
$\lambda_l=\lambda_{1k}(z_l)$, $r'_l=r'|_{\eta'=\eta_l}$. Here
\begin{equation}
z_l=\frac{\nu_{1k}}{\nu}-1,\quad \frac{\nu_{1k}}{1+z_0}\leq \nu\leq\nu_{1k},
\label{zl}
\end{equation}
where $\nu$ is a radiation frequency in the present-day epoch. By its physical
sense $z_l$ is a resonance redshift at which photons with the frequency $\nu$
were scattered and $\tau_l$ is the Sobolev optical thickness for this redshift.
These quantities appear in the theory of primordial hydrogen recombination lines
formation (see e.g., Dubrovich and Grachev, 2004). Further, in eq. (\ref{nexp0})
there is
\begin{equation}
r'=\sqrt{r^2-2r\mu(\eta_0-\eta')+(\eta_0-\eta')^2},
\label{rhomup}
\end{equation}
where $r$ is a distance from the burst center, $\arccos \mu$ is an angular
distance from the direction on the burst center, $\eta$ is a conformal time:
\begin{equation}
\eta=c\int_0^{t}dt'/a(t')=c\int_z^{z_0}dz'/H(z'),
\label{etaz}
\end{equation}
where $a(t)$ is a scale factor, $t$ is a time, $a=1/(1+z)$ is a dimensionless
time
\begin{equation}
u=c\sigma_e\int_0^t n_e(t')dt'=
c\sigma_e\int_z^{z_0} \frac{n_e(z')}{(1+z')H(z')}dz'
\label{udef}
\end{equation}
which has a sence of an optical distance (for Thomson scattering) between
the moments $z$ and $z_0$, $\sigma_e=6.65\cdot 10^{-25}$ cm$^2$ is the Thomson
scattering crossection. Here redshift $z_0$ corresponds to initial moment of
time (a burst moment): $u=\eta=t=0$ for $z=z_0$.

As to $n^0$ we assume that at the initial moment of time $t=0$ ($\eta=0$,
$z=z_0$) the burst radiation is isotropical and spherically symmetrical:
\begin{equation}
n^0(r,\mu,0)=n_0(r),
\end{equation}
where $n_0(r)$ is a given function which we take in the form
\begin{equation}
n_0(r)=\pi^{-3/2}r_*^{-3}\exp[-(r/r_*)^2]\rightarrow \delta(r)/
4\pi\,r^2\quad \mbox{for}\quad r_*\rightarrow 0,
\label{n0rho}
\end{equation}
where $r_*$ is a parameter defining characteristic size of the burst.

\begin{center}
{\bf Main results}
\end{center}

By using the code developed in previous our paper (Dubrovich and Grachev, 2015)
we calculate summary profiles of hydrogen L$_\alpha$ and L$_\beta$ lines and
HeI lines 1s2-1s2p and 1s2-1s3p for three initial redshifts $z_0$: 4000 and
5000 for hydrogen and 6000 for helium.

For the width of initial intensity distribution as a function of $r$ (see
eq. (\ref{n0rho})) we use $r_* = 50$ Mpc in a distance scale at $z=0$. But in
a distance scale corresponding to the burst moment (at $z=z_0$) the width of
initial distribution (for $z_0\gg 1$) turns out to be significantly lower:
$a(z_0)r_*=r_*/(1+z_0)$.

Figure 3 shows spatial distributions of a mean radiation intensity and Figure 4
shows angular distributions of intensity as a result of multiple scatterings on
free electrons. Computations of these distributions were fulfilled using the
code developed by us earlier (Grachev and Dubrovich, 2011). Summary line
profiles are shown in Figures 5 -- 7. They are in frequency regions for which
the Universe is only weakly opaque in the corresponding lines (see Fig. 1 and
also above in the text). It turned out that the profiles in resonance lines do
not depend both on the distance from the burst center and on direction in
contradistinction to the profiles in subordinate lines considered by us earlier
(Dubrovich, Grachev, 2015). The jumps in the profiles
are on the frequencies  $\nu=\nu_{1k}/(1+z_0)$ corresponding to the moment
of appearance of the burst radiation absorption in the lines with laboratory
frequencies $\nu_{1k}$. For hydrogen lines L$_\alpha$ and L$_\beta$
$\nu=$ 616 and 730 GHz for $z_0=4000$ and $\nu=$ 493 and 584 GHz for $z_0=5000$.
For corresponding HeI lines $\nu=$ 855 and 930 GHz for $z_0=6000$. Relative
magnitude of the jumps amounts to $10^{-4}$ --- $3\cdot 10^{-3}$.

It should be marked essential distinction of spectrum distortions profiles for
resonance and subordinate lines. The first ones are purely in absorption and
do not depend both on distance and direction. The second ones may contain also
appreciable emission components and relative contribution of emission and
absorption components noticeably depends both on distance from the source and
on direction (see Dubrovich and Grachev, 2015). The thing is that optical
thickness of the Universe in subordinate lines has quite narrow maximum over
$z$ with subsequent fast decrease but optical thickness in resonance lines
grows nearly exponentially with $z$ decrease (see Fig. 1). Moreover, optical
thickness in all subordinate lines is quite small ($<10^{-3}$) and one
may use linear approximation over optical thickness. But an optical thickness
in resonance lines can become very large so that spectrum calculations becomes
very complicated. However it is clear that in the case of very large optical
thickness a relative amplitude of spectrum distortions may be of the order
unity.

\newpage

\begin{center}
{\bf Conclusion}
\end{center}

The local bursts model of primordial plasma and radiation fluctuations in the
early Universe is considered. These fluctuations may be represented as the local
fastly variable sources of initial blackbody radiation with the temperature
which is slightly different from the average CMB temperature. More generally,
any really detached object (e.g. primordial accreting black hole) may come out
as a source. In this work we calculate transition function from the radiation
intensity of these sources to the observed intensity with the account of photons
scattering on free electrons and in resonance lines of hydrogen and helium. In
the first approximation this function has a physical sense of an optical
thickness for scattering in lines.

As a model we consider scattering of continous radiation of a source
(instantaneous burst of radiation at a given $z_0$) on free electrons and on
atoms of primordial hydrogen in L$_\alpha$ and L$_\beta$ lines and on atoms
of primordial helium (HeI) in lines 1s2-1s2p ($^1$S -- $^1$P*) and 1s2-1s3p
($^1$S -- $^1$P*) at the epoch before hydrogen and He I recombination. For
example, we consider three values of $z_0$: 4000, 5000 and 6000, which are in a
region where optical thickness in the mentioned lines is less than unity. It is
shown that thus arising lines in the source spectrum are in absorption with the
jumps of the value from $10^{-4}$ to $3\cdot 10^{-5}$ and their shape do not
depend both on a distance from a source and on angular distance from direction
to the source center. It should be noted that in a redshifts region where
optical thickness in considered resonance lines is greater than unity a
relative depth of absorption can reach a value of the order unity.
Real observations give maximum amplitude of temperature deviations in
spots about 300 $\mu$K (Adam et al., 2015b). So that the relative magnitude
of temperature fluctuations is in the limits
$\delta T/T\approx 10^{-4}-10^{-5}$ which is more than 1000 times greater than
the value predicted by subordinate lines.

Obtained results are very important for estimation of the prospects to reveal
primordial protoobjects on redshifts up to 5000. One can search such objects
by using the same methods as in searching quasars with great redshifts
i.e. by searching Lyman jumps. So it is necessary to obtain a map of
sufficiently large part of sky with a medium spectral resolution
($\Delta\lambda/\lambda\approx 1-2$\%) in submillimeter range of wavelengths.

This work was supported in part by St. Petersburg State University grant No.
6.38.18.2014.

\begin{center}
{\bf References}
\end{center}

\begin{enumerate}
\item R. Adam, P.A.R. Ade, et al., Planck 2015 results. I. Overview
of products and scientific results, Planck collaboration, arXiv: 1502.01582v1
[astro-ph.CO] (2015a).
\item R. Adam, P.A.R. Ade, et al., Planck 2015 results. IX. Diffuse
component separation: CMB maps, Planck collaboration, arXiv: 1502.05956v1
[astro-ph.CO] (2015b).
\item S.I. Grachev, V.K. Dubrovich, Astrophysics {\bf 34}, 124 (1991).
\item S.I. Grachev, V.K. Dubrovich, Astron. Lett. {\bf 37}, 293 (2011).
\item V.K. Dubrovich,
Astron. Lett. {\bf 29}, 6 (2003).
\item V.K. Dubrovich, S.I. Grachev, Astron. Lett. {\bf 30}, 657 (2004).
\item V.K. Dubrovich, S.I. Glazyrin, Cosmological
dinosaurs, arXiv:1208.3999v1 [astr-ph.CO] 20 Aug (2012).
\item V.K. Dubrovich, S.I. Grachev, Astron. Lett. {\bf 41}, 537 (2015).
\item Yu. Ralchenko, A.E. Kramida, J. Reader and NIST ASD TEAM
(2011). NIST Atomic Spectra Database (ver. 4.1.0). Available: 
http://physics.nist.gov/asd3[2011, September 22]. National Institute of
Standards and Technology, Gaithersburg, MD.
\item D.G. Hummer, P.J. Storey, Mon. Not. R. Astron. Soc.,
{\bf 297}, 1073 (1998).
\end{enumerate}

\newpage
\begin{figure}[p]
\centering

\resizebox{1.0\textwidth}{!}{\includegraphics{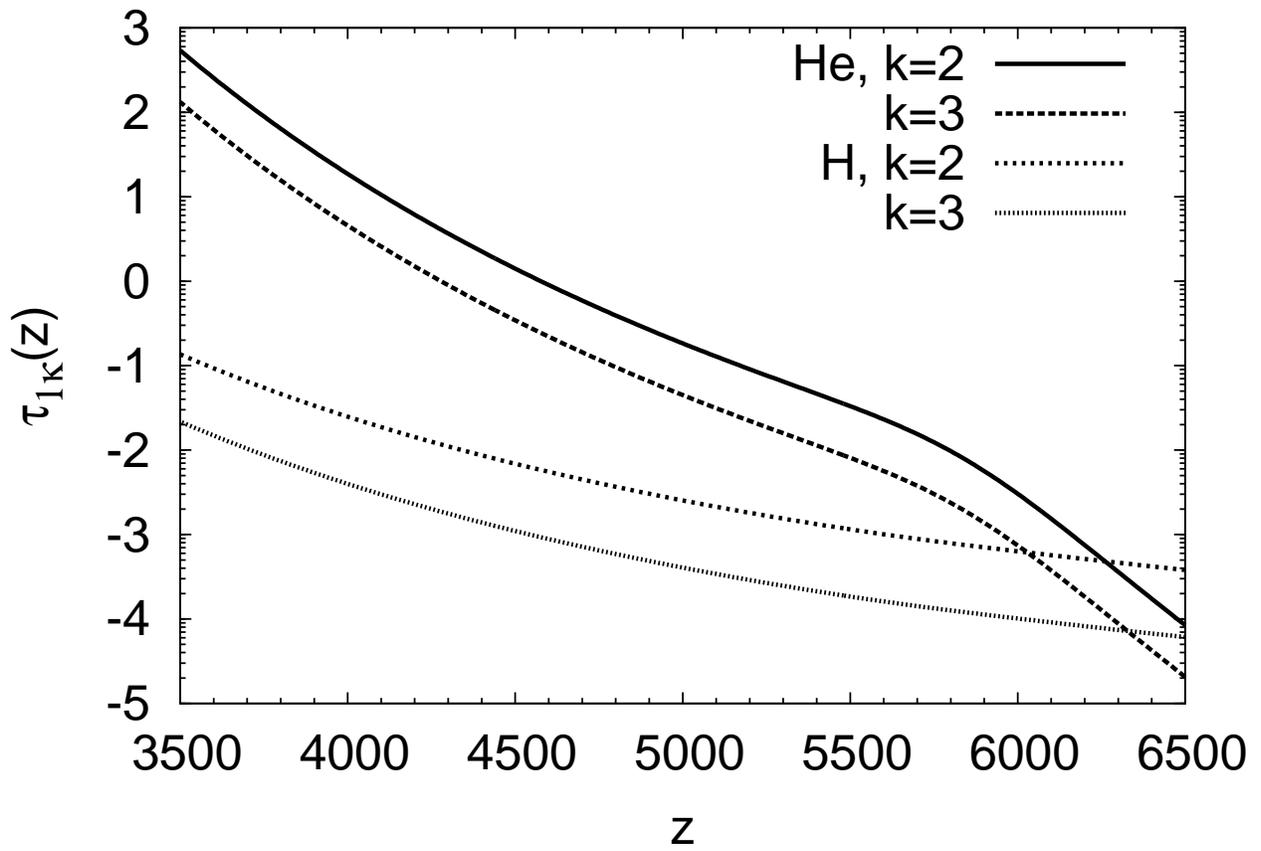}}

\caption {Optical thicknesses in hydrogen L$_\alpha$ and L$_\beta$
lines and in HeI lines 1s2-1skp ($^1$S -- $^1$P*) (k=2 and 3).}

\end{figure}

\begin{figure}[p]
\centering

\resizebox{1.0\textwidth}{!}{\includegraphics{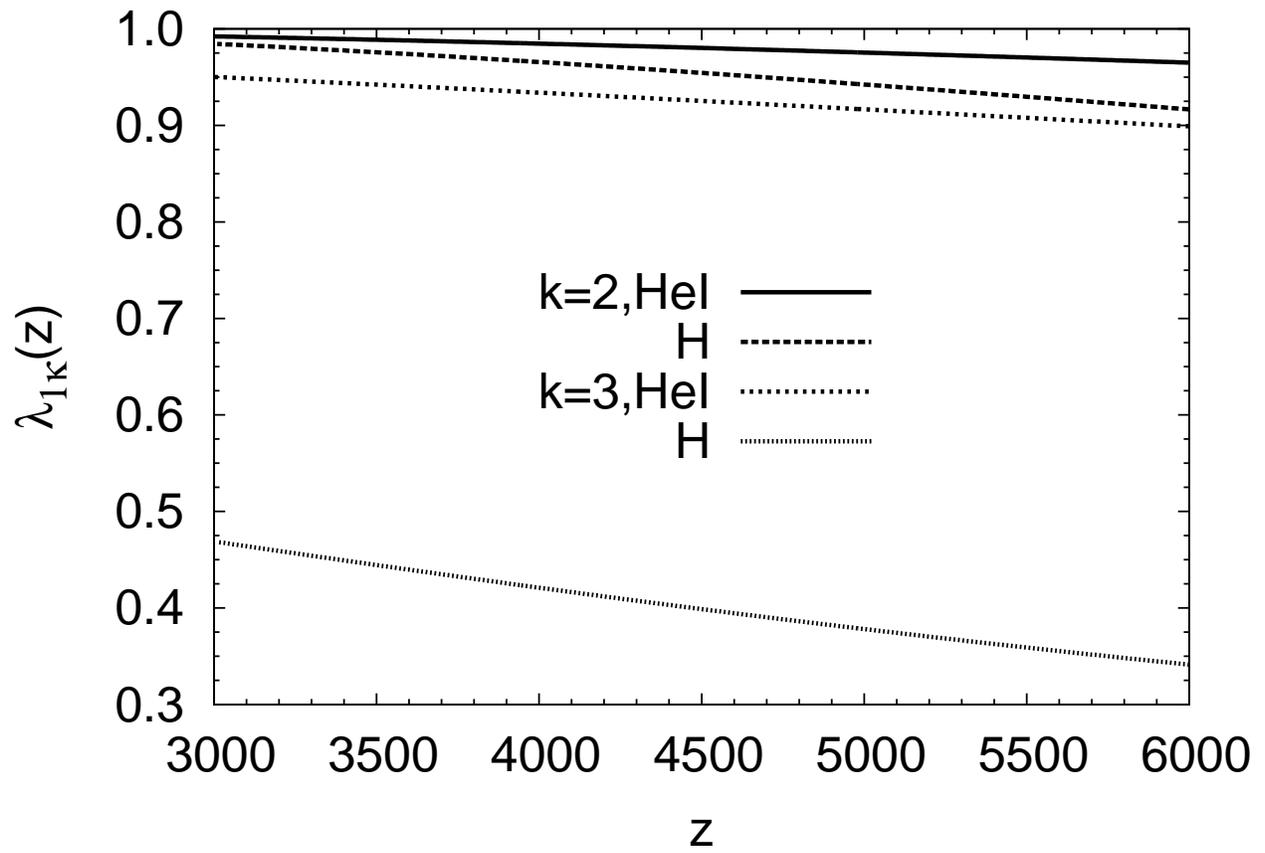}}

\caption{Albedo of a single scattering in the same lines as in the preceeding
figure.}

\end{figure}

\begin{figure}[p]
\centering

\resizebox{1.0\textwidth}{!}{\includegraphics{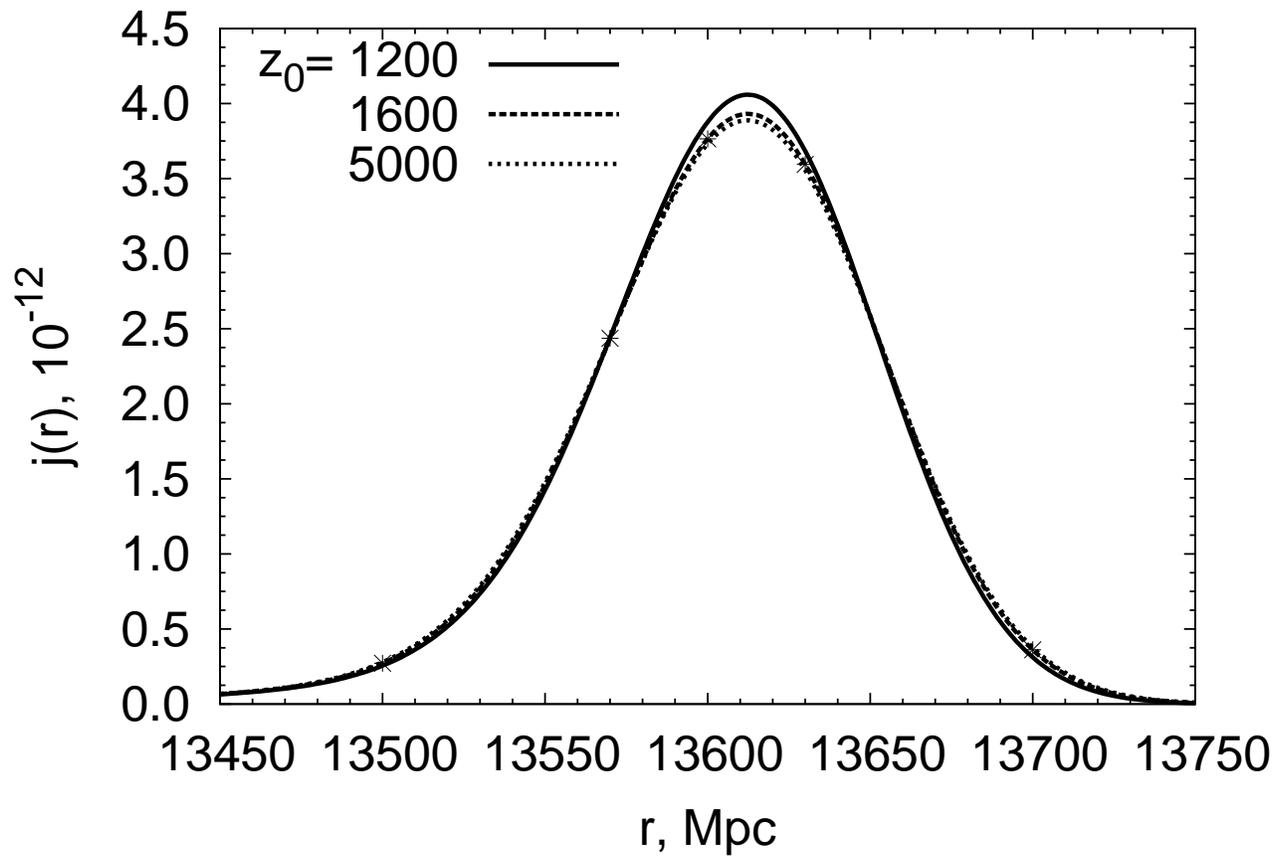}}

\caption{Mean intensity of scattered radiation for bursts at different
$z_0$ (in Mpc). Crosses mark the points in which line profiles were
calculated.}

\end{figure}

\begin{figure}[p]
\centering

\resizebox{1.0\textwidth}{!}{\includegraphics{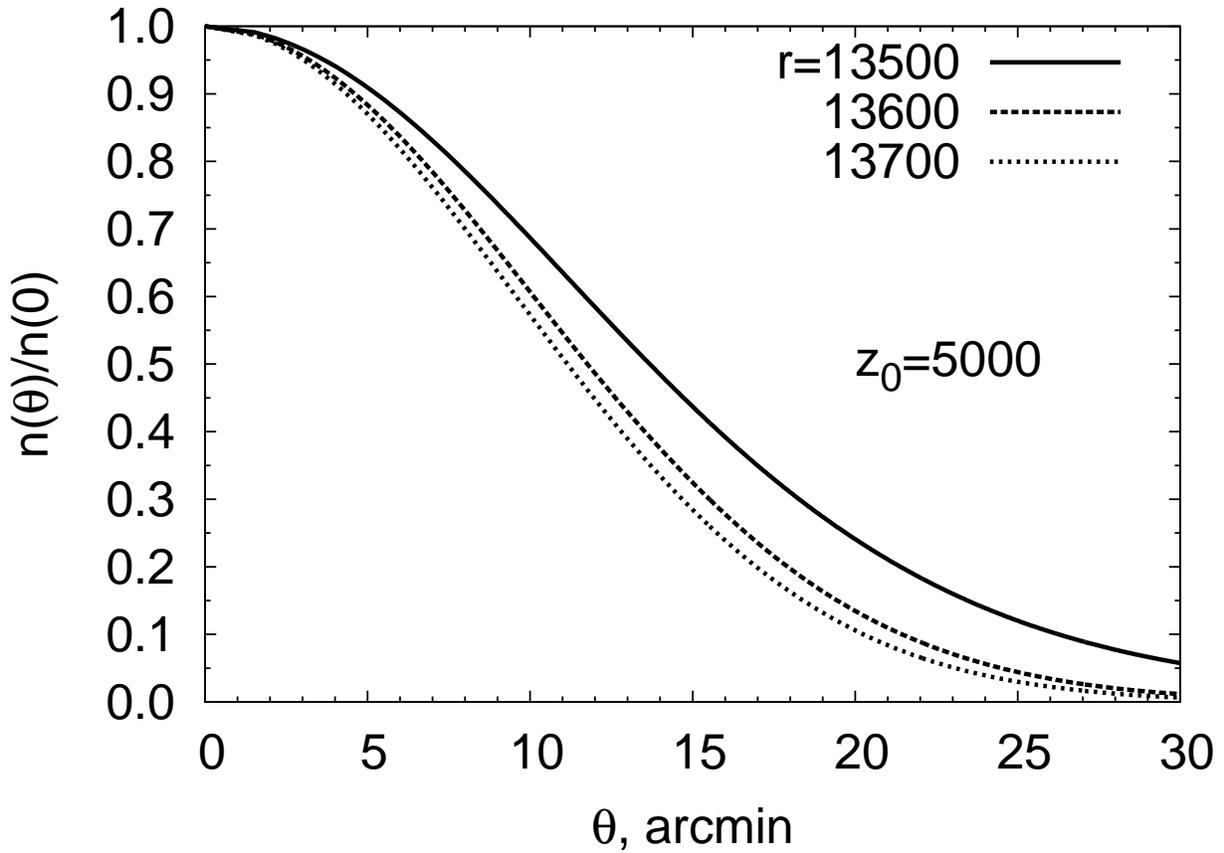}}

\caption{Angular distributions of radiation on different distances $r$ (in Mpc)
from the burst center at the present-day epoch. Here $\theta$ is the
angular distance from the direction on the burst center. In direction on the
burst center $n(0)=4.3\cdot 10^{-8}$, $8.8\cdot
10^{-7}$ and $1.0\cdot 10^{-7}$ for $r=13500$, 13600 and 13700 Mpc respectively.
}

\end{figure}

\begin{figure}[p]
\centering

\resizebox{1.0\textwidth}{!}{\includegraphics{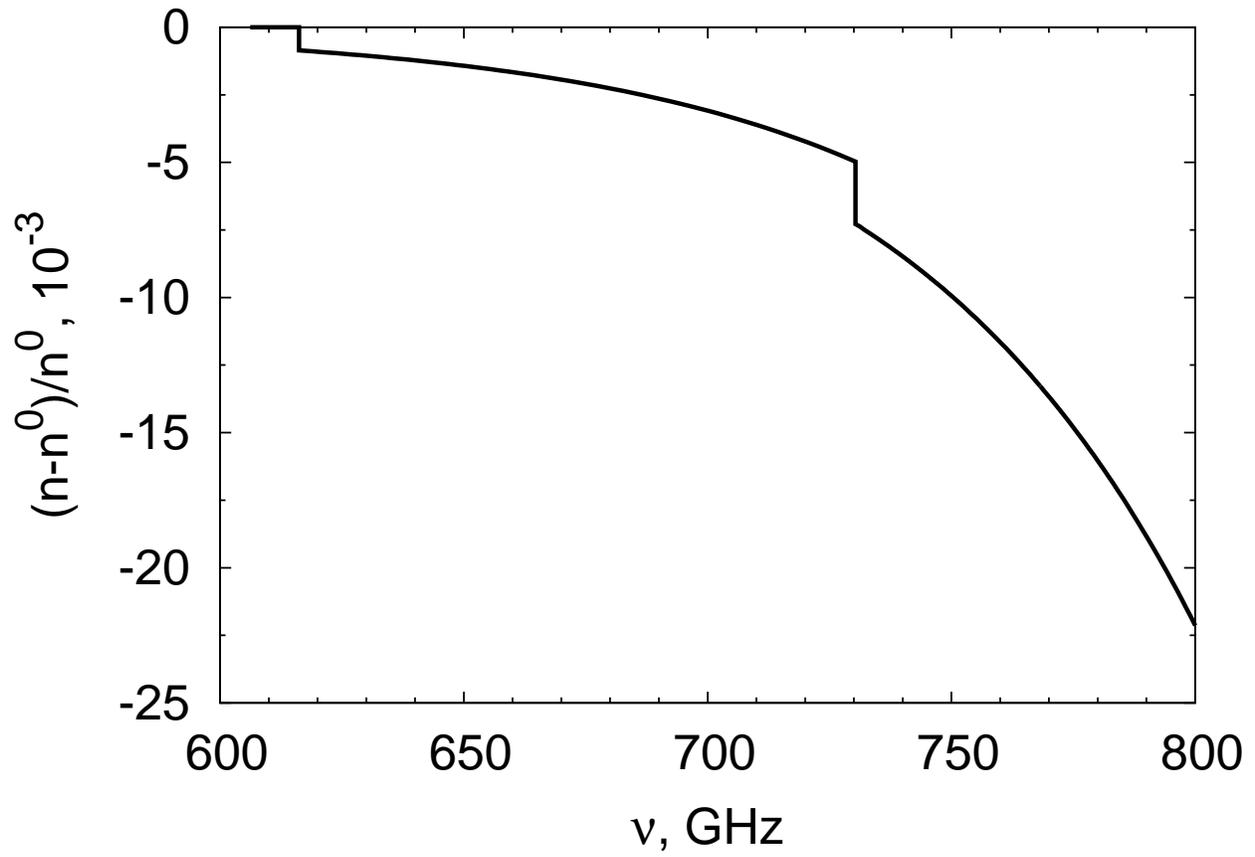}}

\caption{Summary profile of hydrogen lines L$_\alpha$ and L$_\beta$
for the burst at $z_0=4000$.}

\end{figure}

\begin{figure}[p]
\centering

\resizebox{1.0\textwidth}{!}{\includegraphics{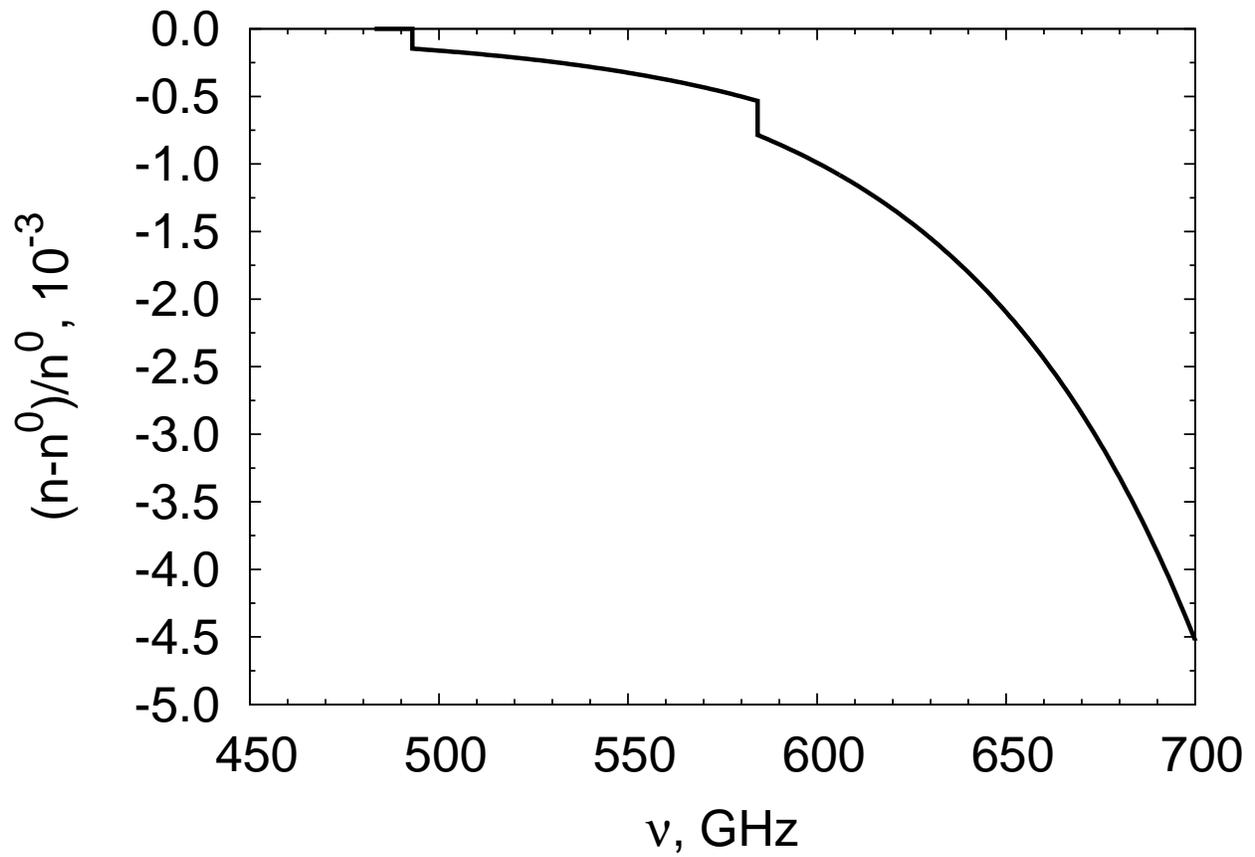}}

\caption {The same as in the preceeding figure but for $z_0=5000$.}

\end{figure}

\begin{figure}[p]
\centering

\resizebox{1.0\textwidth}{!}{\includegraphics{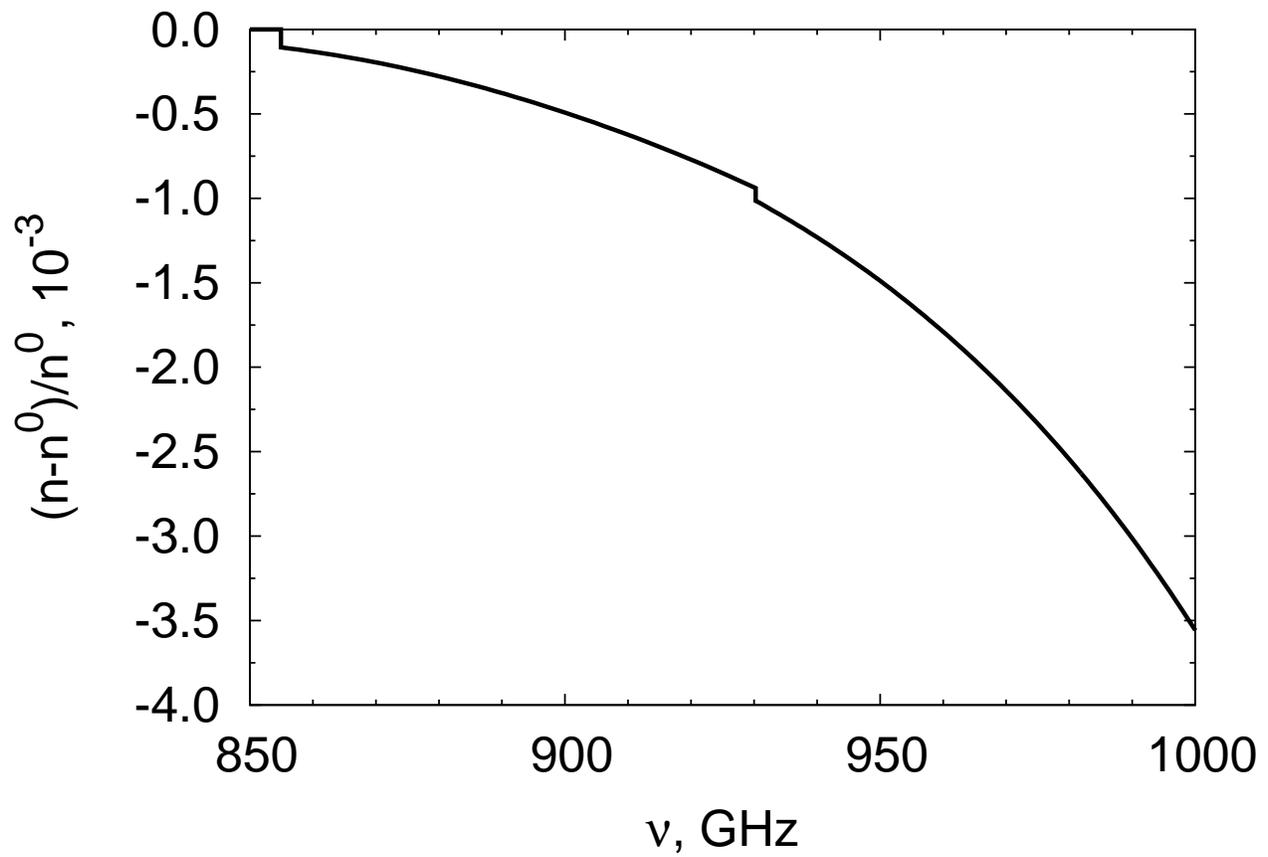}}

\caption{Summary profile of HeI lines 1s2-1s2p ($^1$S -- $^1$P*) and 1s2-1s3p
($^1$S -- $^1$P*) for the burst at $z_0=6000$.}

\end{figure}

\end{document}